\documentclass[11pt,a4paper]{article}
\usepackage[margin=2.54cm]{geometry}
\usepackage{graphicx}
\usepackage{listings}
\usepackage{xcolor}
\usepackage{booktabs}
\usepackage{float}
\usepackage{amsmath}
\usepackage{array}
\usepackage{tabularx}
\usepackage{url}
\usepackage[hidelinks]{hyperref}

\lstdefinestyle{CStyle}{
    language=C,
    basicstyle=\ttfamily\footnotesize,
    keywordstyle=\color{blue},
    commentstyle=\color{gray},
    stringstyle=\color{red},
    numbers=left,
    numberstyle=\tiny,
    stepnumber=1,
    breaklines=true,
    breakatwhitespace=false,
    columns=flexible,
    keepspaces=true,
    showstringspaces=false,
    tabsize=2,
    frame=single,
    xleftmargin=1.2em,
    framexleftmargin=1.2em,
    aboveskip=0.6em,
    belowskip=0.6em,
    captionpos=b
}

\usepackage{authblk}

\begin{document}

\title{When to Treeify Hash Table Buckets:\\A Reproducible C Study of List, Hybrid, and Red-Black Tree Chaining}

\author[1]{Georgii Kashintsev\thanks{Correspondence: \texttt{amoshi.mandrakeuser@gmail.com}. ORCID: \url{https://orcid.org/0009-0005-8220-6022}.}}
\affil[1]{Rambler\&Co, Moscow, Russia}

\date{}

\maketitle

\begin{abstract}
\noindent
\textbf{Practitioner summary.} Do not copy Java's threshold of eight alone: when bins grow long, \textbf{hybrid-batch} (convert after load) still walks lists during insert, while \textbf{hybrid-incremental} (convert as soon as a bin hits~$k$) matches always-tree. Prefer hybrid-incremental or always-tree for overloaded bins; reserve hybrid-batch for pure bulk load then query when chains stay short after resize. Hybrid-incremental approximates Java conversion \emph{timing}, not a \texttt{HashMap} port. Lead metrics below are \texttt{strcmp} counts and heap---more stable than long-list wall-clock.

\smallskip\noindent
When individual hash buckets grow long, linked-list separate chaining incurs linear per-bucket cost. We show that \emph{when} conversion runs (hybrid-batch finalize vs.\ hybrid-incremental) dwarfs the choice of threshold~$k$ for C implementers. Using one C separate-chaining API~\cite{rbtree_repo}, we compare policies under uniform-hash FNV (including a fixed-$m$ probe at $\alpha{\approx}122$), forced-bucket chaining stress, and a moderate-load same-API scale run ($\alpha{=}16$). Under stress, list lookup averages ${\sim}31{,}250$ comparisons vs ${\sim}15$ once treeified; mid-load probes need ${\sim}37$M comparisons under hybrid-batch vs ${\sim}46$k under hybrid-incremental; final post-load comparisons converge (${\sim}15$). Tree buckets use about $1.7\times$ more heap than lists. Stress wall-clock for long lists is illustrative and run-noisy (Appendix~\ref{app:methodology}); we therefore headline comparisons and memory. Replaying real trigram posting-list lengths through the same policies yields the same ranking. At $\alpha{\approx}122$ without resize, some tree wins are really deferred rehash---resize first when $m$ is simply too small.
\end{abstract}


\section{Introduction}
\label{sec:introduction}

Hash tables are ubiquitous in compilers, databases, search systems, and monitoring pipelines. With separate chaining, worst-case per-bucket cost is controlled by the overflow structure---typically a linked list, sometimes a tree after a length threshold as in Java~8+ \texttt{HashMap}~\cite{java_hashmap_treeify}. That Java change was motivated in part by algorithmic complexity attacks (HashDoS): adversarial keys that force long collision chains~\cite{crosby_wallach_hashdos,java8_hashmap_notes}.

\paragraph{Research question.}
For C implementers who control bucket structure but lack a JVM runtime: \emph{when does RB-tree chaining (including hybrid treeification) pay off relative to list chaining, and does conversion \emph{timing}---hybrid-batch versus hybrid-incremental---matter as much as the threshold~$k$?}

We answer with one separate-chaining API, open benchmarks~\cite{rbtree_repo}, and design guidelines---not a new asymptotic result. Section~\ref{sec:chaining-benchmark} isolates policy under uniform-hash and forced-bucket stress; Section~\ref{sec:library-benchmark} gives a same-API scale baseline; Section~\ref{sec:posting-replay} transfers results onto empirical trigram posting lengths.

\paragraph{Contributions.}
\begin{itemize}
\item batch vs.\ incremental hybrid treeification are not interchangeable when copying Java's $k{=}8$ (Section~\ref{sec:chaining-benchmark});
\item a reusable C \texttt{chain\_ht} API with list, hybrid-batch, hybrid-incremental, and always-tree modes;
\item microbenchmarks and a load-factor contrast ($\alpha{\approx}122$ fixed~$m$ vs $\alpha{=}16$ scale);
\item posting-list length CDF plus chain-policy replay on real dictionary trigrams;
\item guidelines coupling policy to per-bin overload and resize vs.\ treeify (Section~\ref{sec:design-guidelines}).
\end{itemize}

Section~\ref{sec:background}--\ref{sec:implementation} give background, related work, and code layout; Section~\ref{sec:evaluation} reports experiments; Sections~\ref{sec:discussion}--\ref{sec:design-guidelines} synthesize advice. Appendix~\ref{app:methodology} documents the measurement protocol.

\section{Background}
\label{sec:background}

A hash table stores key--value pairs in an array of $m$ buckets. A hash function $h$ maps each key to an integer; the bucket index is typically $h(k) \bmod m$, often implemented as a bit mask when $m$ is a power of two~\cite{clrs}. Expected-time lookup is $O(1)$ only if buckets stay short.

\emph{Separate chaining} attaches a secondary structure to each bucket for keys that collide on the index. Linked lists are the default in many libraries; Java~8+ \texttt{HashMap} converts a long bin to a red-black tree when the chain exceeds eight nodes~\cite{java_hashmap_treeify}. Within a bucket of size $N_j$, list lookup costs $\Theta(N_j)$ key comparisons in the worst case, whereas a balanced tree costs $O(\log N_j)$~\cite{clrs}.

This work evaluates \emph{RB-tree chaining}: each non-empty bucket holds an RB tree ordered by key~\cite{clrs}. The question is not whether trees can replace lists in theory---that is well known---but when the extra maintenance pays off under uniform-hash keys at controlled load factor, deliberately overloaded bins, and empirically long posting lists, relative to list chaining, hybrid treeification, and existing library tables.

\section{Related Work}
\label{sec:related}

\subsection{Separate chaining and bucket overflow structures}

Separate chaining is the classical strategy analyzed in Knuth~\cite{knuth_taocp_v1} and in modern algorithms texts~\cite{clrs}. If $n$ elements are stored in a table of size $m$ and bucket $j$ contains $N_j$ elements, then a successful search in a list-based bucket requires $\Theta(N_j)$ key comparisons in the worst case for that bucket, while average behavior depends on the load factor $\alpha = n/m$ and on the hash-function distribution.

When bucket chains grow long, two well-known responses appear in both theory and practice: resize/rehash the table, or replace the bucket representation with a structure that reduces lookup cost within the bucket. The present paper focuses on the second option and compares list chaining with RB-tree chaining under controlled and application-driven workloads.

\subsection{Industrial hash tables with tree buckets}

The most widely cited production example is \texttt{java.util.HashMap} in Java~8 and later. For keys that implement \texttt{Comparable}, Java converts an overloaded bin from a linked list into a red-black tree once the chain exceeds a threshold (8 nodes before treeification, 6 nodes when converting back)~\cite{java_hashmap_treeify,java8_hashmap_notes}. A principal motivation is worst-case degradation from hash collisions---including HashDoS-style adversarial key streams that force long chains under weak or predictable hashing~\cite{crosby_wallach_hashdos}---while preserving list-based behaviour for short bins where tree overhead would not pay off.

Other mainstream standard-library hash tables take different trade-offs. For example, many C++ \texttt{unordered\_map} implementations remain list-based buckets~\cite{cpp_unordered_map}, Python dictionaries use open addressing rather than tree buckets~\cite{python_dict_design}, and high-performance C libraries such as TommyDS provide list-based dynamic chaining~\cite{tommyds}. These choices show that tree buckets are an established but not universal design point.

\subsection{Ordered and searchable bucket variants}

Tree buckets also relate to ordered associative containers that keep keys sorted within nodes or buckets~\cite{google_cpp_btree,pugh_skip_list}. Those designs target range access or ranked queries; here we study overflow structure inside a hash table whose primary operation remains exact key lookup under skewed bucket load.

\subsection{Position of this work}

Prior work already establishes that tree-based buckets can mitigate long-chain worst cases. What is less often reported for C implementers is a reproducible comparison that separates (i)~list versus hybrid versus always-tree in one API, (ii)~\emph{batch} versus \emph{incremental} hybrid conversion timing, and (iii)~transfer of that policy result onto real posting-list length distributions (Sections~\ref{sec:chaining-benchmark}--\ref{sec:skewed-applications}). This paper therefore contributes an empirical positioning study and an open benchmark suite, rather than a new asymptotic result. The finding that treeification \emph{policy} matters as much as the threshold~$k$ complements Java's published motivation for treeifying long bins~\cite{java8_hashmap_notes}.

\section{Implementation}
\label{sec:implementation}

All experiments share a modular C codebase published with this paper~\cite{rbtree_repo}.

\paragraph{Chaining comparison API.} \texttt{chain\_ht.c} exposes list, always-tree, and hybrid bucket modes with either \textbf{batch} or \textbf{incremental} treeify policy behind the same insert/lookup interface (\texttt{chaining\_benchmark}; Section~\ref{sec:chaining-benchmark}). Supporting drivers for trigram posting-list statistics (\texttt{inverted\_chain\_bench}) and optional trie/quantile experiments ship in the repository~\cite{rbtree_repo}.

\section{Evaluation}
\label{sec:evaluation}

We test the research question in three stages. Section~\ref{sec:chaining-benchmark} holds the \emph{controlled} evidence: the same chaining API under uniform FNV load and a forced-bucket chaining stress, including batch versus incremental hybrid policy. Section~\ref{sec:library-benchmark} reports a same-API in-memory scale baseline. Section~\ref{sec:skewed-applications} transfers the policy result onto empirical trigram posting-list lengths.

\subsection{List versus tree bucket chaining}
\label{sec:chaining-benchmark}

To isolate \emph{bucket overflow structure} from hash-function quality, we implemented list, hybrid, and always-tree chaining behind the same API (\texttt{chain\_ht}) and measured them with \texttt{chaining\_benchmark}~\cite{rbtree_repo}. The research object is the secondary structure inside a bin---not a claim about how often FNV produces Zipf collisions over an entire table.

\paragraph{Uniform-hash workload (fixed table; high load factor).}
Inserts 500\,000 unique string keys into a \emph{fixed} table of 4096 buckets using FNV (no resize). Keys are hashed uniformly in the usual sense, but mean load factor is $\alpha = n/m \approx 122$, so chains are not short: list avg.\ comparisons ${\sim}62$ and max chain 167 (Table~\ref{tab:chaining-uniform}). Tree/hybrid benefit on this row is therefore partly a \emph{fixed-size} effect---a production table would usually rehash long before $\alpha{\approx}122$. We keep the configuration to isolate bucket overflow structure at known $\alpha$; the fairer ``moderate uniform'' comparison is the same-API scale baseline at $\alpha{=}16$ (Table~\ref{tab:scale-baseline}).

\paragraph{Structural chaining stress (forced-bucket overload).}
Uses the same $N$ and $m$, but keys embed a bucket id in the prefix and the table hash returns that id, concentrating keys into eight hot bins of ${\sim}62{,}500$ entries each. This is \emph{not} hash skew in the Zipfian sense: it deliberately bypasses the hash so that chain length is the independent variable. The point is a per-bucket stress probe of list scan, batch treeify, and incremental treeify---the same failure mode long posting lists share ($\Theta(L)$ work inside one bin), not a model of ``eight equal megachains as a realistic $n$-gram hash mix.'' Realism of \emph{lengths} is deferred to Section~\ref{sec:posting-replay}.

\paragraph{Two hybrid policies (not one ``hybrid'').}
\label{para:hybrid-protocol}
We never treat bare ``hybrid'' as a complete policy. Hybrid mode shares the familiar threshold~$k$ (default eight, after Java~\cite{java_hashmap_treeify}), but \emph{conversion timing} differs:

\begin{itemize}
\item \textbf{Hybrid-batch:} inserts always go into lists; \texttt{chain\_ht\_finalize} converts eligible bins after the load phase. This is a ``treeify at the end of bulk load'' engineering shortcut---\emph{not} Java behaviour.
\item \textbf{Hybrid-incremental:} a bin converts to an RB tree \emph{as soon as} its chain reaches~$k$ during insert. This approximates Java's conversion \emph{timing} under insert/lookup interleaving, not a reimplementation of \texttt{java.util.HashMap} (we omit untreeify hysteresis, resize packaging, and \texttt{Comparable} tree-bin details).
\end{itemize}

\paragraph{Protocol note.}
Copying only the constant~$k$ while using hybrid-batch does \emph{not} reproduce Java's policy and can make post-load lookup look excellent while understating online insert/mid-load cost. That distinction is the central practical finding of this paper. Guidelines and tables always spell batch vs.\ incremental.

Unlike \texttt{java.util.HashMap}, which treeifies at eight nodes and untreeifies below six~\cite{java_hashmap_treeify,java8_hashmap_notes}, some repository drivers (legacy string table) attach an RB tree once a bucket becomes non-empty. The reason is not that trees always win on paper for tiny buckets: for chain length~$n$, a list lookup needs up to $n$ key comparisons in the worst case, whereas a balanced tree needs at most about $\lfloor\log_2 n\rfloor+1$ on a successful search~\cite{clrs}. The gap is small for the chain lengths Java optimizes for:
\begin{table}[H]
\centering
\caption{Successful-search key comparisons within one bucket (worst case, list vs.\ balanced RB tree).}
\label{tab:chain-comparisons}
\begin{tabular}{@{}rrr@{}}
\toprule
Chain length & List & RB tree \\
\midrule
8 & 8 & 3 \\
7 & 7 & 3 \\
6 & 6 & 3 \\
5 & 5 & 3 \\
4 & 4 & 2 \\
3 & 3 & 2 \\
2 & 2 & 1 \\
1 & 1 & 1 \\
\bottomrule
\end{tabular}
\end{table}
Therefore Java's threshold policy is reasonable when most bins stay below eight elements. Our always-tree choice trades slightly higher insertion overhead on sparse buckets for simpler code paths and predictable behavior when bucket load is unknown a priori---as in inverted-index posting lists that can grow from one to thousands of entries under the same n-gram key.

\paragraph{Batch versus incremental under bulk load.}
Table~\ref{tab:chaining-policy} reports single-run measurements (seed~1, 2026-07-15). Primary outcome is avg.\ comparisons: under forced-bucket stress, post-load lookup converges to ${\sim}15.3$ for batch hybrid, incremental hybrid, and always-tree---so measuring only the final search phase hides the batch tax. That tax shows up in mid-load probes (Table~\ref{tab:chaining-mix}: ${\sim}37$M vs ${\sim}46$k comparisons) and in wall-clock during insert; wall-clock for long-list phases is illustrative and run-noisy (Appendix~\ref{app:methodology}), whereas comparison counts with a fixed seed are deterministic.

\begin{table}[H]
\centering
\caption{Batch vs.\ incremental hybrid ($k{=}8$) against list and always-tree baselines ($5\times10^5$ keys, 4096 buckets, seed~1). Prefer the comparison column; wall times for long lists are single-run and noisy. ``Stress'' = forced-bucket overload, not Zipf hash skew.}
\label{tab:chaining-policy}
\begin{tabular}{@{}llrrr@{}}
\toprule
Workload & Policy & Insert (s) & Lookup (s) & Avg.\ comparisons \\
\midrule
Uniform & List & 0.84 & 0.81 & 62.1 \\
Uniform & Hybrid-batch & 0.91 & 0.060 & 6.5 \\
Uniform & Hybrid-incremental & 0.14 & 0.055 & 6.4 \\
Uniform & Always RB tree & 0.14 & 0.051 & 6.5 \\
\midrule
Stress & Hybrid-batch & 121.2 & 0.040 & 15.3 \\
Stress & Hybrid-incremental & 0.20 & 0.044 & 15.3 \\
Stress & Always RB tree & 0.20 & 0.043 & 15.3 \\
\bottomrule
\end{tabular}
\end{table}

\paragraph{Interleaved mid-load lookups.}
Table~\ref{tab:chaining-mix} adds probes during insertion (every 10{,}000 inserts, 64 random earlier keys). Under hybrid-batch those probes still walk growing lists (${\sim}37$M comparisons); hybrid-incremental already holds trees (${\sim}46$k comparisons). This is closer to production ``bulk load with occasional online traffic'' than a pure insert-then-search microbenchmark---and is why hybrid-batch must not be read as a Java-\texttt{HashMap} surrogate.

\begin{table}[H]
\centering
\caption{Forced-bucket stress with interleaved mid-load lookups ($\texttt{mix\_every}{=}10{,}000$, 64 probes; $k{=}8$). Hybrid-batch vs hybrid-incremental only.}
\label{tab:chaining-mix}
\begin{tabular}{@{}lrrr@{}}
\toprule
Policy & Insert+mid (s) & Mid comparisons & Final lookup (s) \\
\midrule
Hybrid-batch & 93.1 & $36{,}898{,}600$ & 0.040 \\
Hybrid-incremental & 0.19 & $46{,}275$ & 0.042 \\
\bottomrule
\end{tabular}
\end{table}

For completeness, Table~\ref{tab:chaining-uniform} reports five-run means from \textbf{uniform-only} repetitions (no forced-bucket runs between repetitions; Appendix~\ref{app:methodology}). Table~\ref{tab:chaining-skew} isolates forced-bucket stress: the stable signal is avg.\ comparisons (${\sim}31{,}250$ list vs ${\sim}15$ tree/hybrid), not the absolute list wall-clock. Table~\ref{tab:k-sweep} sweeps hybrid-incremental $k{=}1..8$ on the uniform-hash fixed-$m$ load: every bin eventually treeifies (4096 events), so final avg.\ comparisons stay ${\sim}6.4$--$6.5$. That supports the claim that conversion \emph{timing}---not the particular $k\in[1,8]$---dominates once bins exceed the threshold early in the load. On forced-bucket stress the same pattern holds for final post-load comparisons once conversion completes.

\begin{table}[H]
\centering
\caption{Hybrid-incremental threshold sweep on uniform-hash fixed table ($5\times10^5$ keys, 4096 buckets, $\alpha{\approx}122$, seed~1; single run each~$k$).}
\label{tab:k-sweep}
\begin{tabular}{@{}rrrrr@{}}
\toprule
$k$ & Insert (s) & Lookup (s) & Avg.\ comparisons & Treeify events \\
\midrule
1 & 0.165 & 0.054 & 6.45 & 4096 \\
2 & 0.169 & 0.055 & 6.45 & 4096 \\
3 & 0.172 & 0.056 & 6.45 & 4096 \\
4 & 0.170 & 0.056 & 6.44 & 4096 \\
5 & 0.166 & 0.054 & 6.43 & 4096 \\
6 & 0.164 & 0.057 & 6.44 & 4096 \\
7 & 0.161 & 0.056 & 6.43 & 4096 \\
8 & 0.181 & 0.055 & 6.44 & 4096 \\
\bottomrule
\end{tabular}
\end{table}

\begin{table}[H]
\centering
\caption{Chaining microbenchmark, uniform-hash keys on a \emph{fixed} table ($5\times10^5$ keys, 4096 buckets, $\alpha{\approx}122$, FNV, seed~1). Insert and lookup: mean wall time (s) over five \emph{uniform-only} runs (2026-07-15). Not ``short chains''---see max chain / avg.\ comparisons.}
\label{tab:chaining-uniform}
\begin{tabular}{@{}lrrrr@{}}
\toprule
Bucket policy & Insert (s) & Lookup (s) & Avg.\ comparisons & Max chain \\
\midrule
List & 1.07 & 1.00 & 62.1 & 167 \\
Hybrid-batch ($k{=}8$) & 1.15 & 0.059 & 6.5 & 167 \\
Hybrid-incremental ($k{=}8$) & 0.17 & 0.055 & 6.4 & 167 \\
Always RB tree & 0.16 & 0.053 & 6.5 & 167 \\
\bottomrule
\end{tabular}
\end{table}

\begin{table}[H]
\centering
\caption{Forced-bucket stress ($5\times10^5$ keys, eight hot bins, seed~1; \textbf{single run}). Headline: avg.\ comparisons (${\sim}31$k$\to$${\sim}15$). Hybrid-batch is \emph{not} Java-like---see hybrid-incremental row and Table~\ref{tab:chaining-mix}. List wall-clock is illustrative only.}
\label{tab:chaining-skew}
\begin{tabular}{@{}lrrrr@{}}
\toprule
Bucket policy & Insert (s)$^\dagger$ & Lookup (s)$^\dagger$ & Avg.\ comparisons & Max chain \\
\midrule
List & 271.5 & 338.3 & 31250.5 & 62500 \\
Hybrid-batch ($k{=}8$) & 121.2 & 0.040 & 15.3 & 62500 \\
Hybrid-incremental ($k{=}8$) & 0.20 & 0.044 & 15.3 & 62500 \\
Always RB tree & 0.20 & 0.043 & 15.3 & 62500 \\
\bottomrule
\end{tabular}\\[0.5em]
\footnotesize $^\dagger$Wall times for overloaded list/hybrid-batch insert phases; do not treat ratios of these seconds as the primary claim.
\end{table}

\begin{table}[!t]
\centering
\footnotesize
\caption{Heap memory after uniform-workload inserts ($5\times10^5$ keys, 4096 buckets, seed~1). \texttt{heap\_bytes} counts bucket arrays, list cells or RB nodes, and stored key strings via \texttt{chain\_ht\_heap\_bytes()}.}
\label{tab:chaining-memory}
\begin{tabular}{@{}lrr@{}}
\toprule
Bucket policy & Heap (MB) & Bytes/key \\
\midrule
List & 23.1 & 46.3 \\
Hybrid-batch / hybrid-incremental after treeify ($k{=}8$) & 39.1 & 78.3 \\
Always RB tree & 39.1 & 78.3 \\
\bottomrule
\end{tabular}
\end{table}

Where bins become long---whether by forced routing, empirical posting lengths (Section~\ref{sec:skewed-applications}), or a deliberately fixed high $\alpha$---tree chaining (especially hybrid-incremental or always-tree) cuts per-bin comparisons despite higher memory use. When $\alpha$ is merely large because the table was not resized (Table~\ref{tab:chaining-uniform}, $\alpha{\approx}122$), rehash is usually the first remedy; treeify is for cases where length stays high after a sane table size. At moderate $\alpha{=}16$ (Table~\ref{tab:scale-baseline}), lists remain competitive on insert while hybrid-incremental/tree still reduce lookup comparisons.

\subsection{Same-API scale baseline (and optional library context)}
\label{sec:library-benchmark}

To isolate bucket policy at larger uniform load without mixing collision strategies or disk I/O, Table~\ref{tab:scale-baseline} compares list, hybrid-batch, hybrid-incremental, and always-tree behind the \emph{same} \texttt{chain\_ht} API on $2^{20}$ FNV keys and $2^{16}$ buckets (in-memory insert then lookup; load factor $\alpha{=}16$). This is the moderate-load counterpart to Table~\ref{tab:chaining-uniform}'s $\alpha{\approx}122$ fixed-size run. Max chain is only 36, so lists remain competitive: tree/hybrid cut average comparisons from 9.0 to ${\sim}3.6$ and lookup time from 0.65\,s to ${\sim}0.16$\,s. Hybrid-incremental again avoids the hybrid-batch convert-at-end tax on insert (0.61\,s vs 1.08\,s).

Table~\ref{tab:load-factor} summarizes that load-factor sensitivity: at $\alpha{\approx}122$ without resize, list avg.\ comparisons are already ${\sim}62$; at $\alpha{=}16$ they fall to 9.0. Treeify still helps both rows, but resizing is the first response when high $\alpha$ is an artefact of too small~$m$.

\begin{table}[H]
\centering
\caption{Load-factor sensitivity (uniform FNV, same \texttt{chain\_ht} API; avg.\ comparisons after load).}
\label{tab:load-factor}
\begin{tabular}{@{}lrrrr@{}}
\toprule
Setting & $\alpha$ & Max chain & List avg.\ cmp & Tree avg.\ cmp \\
\midrule
Fixed $m{=}4096$, $n{=}5\times10^5$ & ${\approx}122$ & 167 & 62.1 & 6.5 \\
Scale $m{=}2^{16}$, $n{=}2^{20}$ & 16 & 36 & 9.0 & 3.6 \\
\bottomrule
\end{tabular}
\end{table}

\begin{table}[H]
\centering
\caption{In-memory same-API scale baseline ($2^{20}$ keys, $2^{16}$ buckets, FNV, seed~1; no disk I/O).}
\label{tab:scale-baseline}
\begin{tabular}{@{}lrrrr@{}}
\toprule
Policy & Insert (s) & Lookup (s) & Avg.\ comparisons & Max chain \\
\midrule
List & 0.89 & 0.65 & 9.0 & 36 \\
Hybrid-batch ($k{=}8$) & 1.08 & 0.16 & 3.7 & 36 \\
Hybrid-incremental ($k{=}8$) & 0.61 & 0.16 & 3.6 & 36 \\
Always RB tree & 0.62 & 0.16 & 3.6 & 36 \\
\bottomrule
\end{tabular}
\end{table}

For context against third-party implementations, Table~\ref{tab:library-benchmark} reports an older single-run comparison of GNU \texttt{hsearch}~\cite{posix_hsearch}, TommyDS \texttt{tommy\_hashdyn}, and our always-tree table on $2^{25}$ keys with an intervening key-file write/read. Those rows compare different collision strategies (open addressing vs.\ list vs.\ always-tree) under a workload that also includes key-file I/O, so they are \emph{not} a clean list-versus-tree contest; they only show that always-tree chaining is not catastrophic versus common libraries under uniform random keys.

\begin{table}[H]
\centering
\caption{External library context ($2^{25}$ uniform keys; single run; phases include key-file I/O). Not an apples-to-apples bucket-policy comparison---see Table~\ref{tab:scale-baseline}.}
\label{tab:library-benchmark}
\begin{tabular}{@{}lrr@{}}
\toprule
Implementation & Insert (s) & Lookup (s) \\
\midrule
GNU \texttt{hsearch} & 16.186 & 12.091 \\
TommyDS \texttt{tommy\_hashdyn} & 13.044 & 8.120 \\
Custom RB-tree table & 12.091 & 9.047 \\
\bottomrule
\end{tabular}
\end{table}
\subsection{Empirical long-chain lengths from an inverted index}
\label{sec:skewed-applications}

The forced-bucket stress in Section~\ref{sec:chaining-benchmark} answers a mechanism question: \emph{given} long bins, how do list / batch-hybrid / incremental-hybrid / tree behave? It does not claim that real hash tables route mass into eight equal bins. We therefore ask a separate transfer question: do real application artifacts already exhibit per-structure lengths $L\gg 8$? For a trigram inverted index of the kind used for approximate string matching over lexicons~\cite{zobel_dart_approx}, those lengths are posting-list sizes.

\subsubsection{Posting-list length CDF and chain replay}
\label{sec:posting-replay}

We built a trigram-only inverted index over 370{,}105 dictionary words from \texttt{words\_alpha.txt}~\cite{words_alpha} (9{,}165 distinct trigrams). Table~\ref{tab:posting-cdf} is the empirical \emph{length} CDF of those posting lists---the real $n$-gram artefact whose long tails motivate treeify. (The production outer map is keyed by trigram; posting length is the overflow structure size we care about for chaining policy. We do not claim a separate Zipf CDF of FNV bucket occupancy over English word hashes.)

\begin{table}[H]
\centering
\caption{Trigram posting-list length distribution on \texttt{words\_alpha.txt} (370{,}105 words, 9{,}165 distinct trigrams). Empirical long-chain sizes for the inverted index.}
\label{tab:posting-cdf}
\begin{tabular}{@{}lrrrrrr@{}}
\toprule
& Min & $p_{50}$ & $p_{90}$ & $p_{99}$ & Max & Mean \\
\midrule
Posting length & 1 & 29 & 793 & 4{,}002 & 23{,}501 & 300.5 \\
\bottomrule
\end{tabular}
\end{table}

To close the list-versus-tree loop on those lengths, we \emph{replay} the hottest 256 posting lengths as synthetic bucket chains in \texttt{chain\_ht} (list, hybrid-batch, hybrid-incremental, and always-tree). We do not insert dictionary words into \texttt{chain\_ht}: for each posting list of length~$L$ we force one bucket and insert $L$ synthetic keys, so the bucket chain length equals that posting length (one bucket per trigram; $1{,}065{,}810$ keys total). This isolates overflow-structure cost under realistic~$L$, not a redesign of the production trigram$\to$list map. Table~\ref{tab:posting-replay} is therefore an inverted-index-length A/B on the shared chaining API (not a claim that the shipping outer HT must be switched from always-tree to list). List lookup averages thousands of comparisons; hybrid-batch still pays list insert cost before conversion; hybrid-incremental matches always-tree.

\begin{table}[H]
\centering
\caption{Chain-policy replay of the hottest 256 trigram posting lengths (1{,}065{,}810 keys; wall time on Apple M4 Pro, seed-independent lengths).}
\label{tab:posting-replay}
\begin{tabular}{@{}lrrr@{}}
\toprule
Policy & Insert (s) & Lookup (s) & Avg.\ comparisons \\
\midrule
List & 6.62 & 6.57 & 2880.7 \\
Hybrid-batch ($k{=}8$) & 7.35 & 0.082 & 11.7 \\
Hybrid-incremental ($k{=}8$) & 0.22 & 0.080 & 11.7 \\
Always RB tree & 0.22 & 0.077 & 11.7 \\
\bottomrule
\end{tabular}
\end{table}

Thus batch-versus-incremental is not an artefact of eight synthetic bins alone: under real dictionary posting lengths, copying only Java's threshold while using hybrid-batch finalize remains the expensive path. Replaying the hottest 2{,}048 lists (2.46M keys) yields the same ordering (hybrid-batch insert 9.2\,s vs hybrid-incremental 0.47\,s).

The lengths in Table~\ref{tab:posting-cdf} come from the repository's trigram inverted-index driver (\texttt{inverted\_chain\_bench} / \texttt{ngram.c}) over \texttt{words\_alpha.txt}. A fuller inverted-index versus prefix-trie query comparison and an ordered-bucket quantile case study ship with the code~\cite{rbtree_repo} but are omitted here to keep the SPE manuscript focused on bucket treeification policy.

\section{Discussion}
\label{sec:discussion}

Evidence splits into two layers. \emph{Layer~A} (Section~\ref{sec:chaining-benchmark}) is a controlled chaining stress: forced routing creates long bins so overflow structure and treeify \emph{policy} are attributable. \emph{Layer~B} (Table~\ref{tab:posting-cdf} and Table~\ref{tab:posting-replay}) shows that Zipf-like long lengths already occur as trigram posting lists, and that replaying those lengths preserves the batch-versus-incremental ordering. Neither layer claims that FNV over English keys routinely collapses into eight equal megachains.

Hybrid treeification is not one algorithm. With \emph{hybrid-batch}, a $k{=}8$ policy under forced-bucket overload still pays list work until finalize (Table~\ref{tab:chaining-mix}: ${\sim}37$M mid-load comparisons). With \emph{hybrid-incremental}, mid-load probes stay near always-tree (${\sim}46$k comparisons) and final lookup comparisons converge (${\sim}15$). Always-tree buckets remain the simplest option when per-bin load is unknown or routinely large; hybrid-incremental is the closest conversion-\emph{timing} stand-in for Java among our policies, while still keeping short bins as lists on uniform loads. Under moderate uniform load the same API shows lists remain competitive while trees still cut lookup comparisons (Table~\ref{tab:scale-baseline}).

\subsection{Design guidelines for implementers}
\label{sec:design-guidelines}

Table~\ref{tab:design-guidelines} summarizes when each bucket policy is appropriate on the evidence in this paper. The decision is driven primarily by \emph{expected per-bin length} and by whether conversion is batch or incremental---never by an ambiguous bare ``hybrid'' label. Hybrid-incremental approximates Java's treeify \emph{when}, not the full \texttt{HashMap} implementation.

\begin{table}[!t]
\centering
\caption{Bucket-policy selection for separate chaining in C (always name batch vs.\ incremental)}
\label{tab:design-guidelines}
\begin{tabular}{@{}p{0.28\textwidth}p{0.66\textwidth}@{}}
\toprule
Policy & Choose when \\
\midrule
Linked list & Chains stay short after a reasonable resize (median $\ll 8$); memory is tight; keys are uniformly hashed. \\
Hybrid-batch ($k{=}8$) & Pure bulk load then query, with moderate chain lengths and \emph{no} mid-load traffic; do \emph{not} treat as Java-\texttt{HashMap} (Tables~\ref{tab:chaining-policy}--\ref{tab:chaining-mix}). \\
Hybrid-incremental ($k{=}8$) & Prefer this over hybrid-batch when bins can grow during load or lookups interleave with inserts; approximates Java conversion timing, not a HashMap port (Tables~\ref{tab:chaining-policy} and~\ref{tab:chaining-mix}). \\
Always RB tree & Per-bin load is large by construction or worst-case lookup latency dominates; matches hybrid-incremental on forced-bucket stress while accepting ${\sim}1.7\times$ heap (Table~\ref{tab:chaining-memory}). \\
Rehash / enlarge & First choice when $\alpha$ is high only because $m$ is fixed too small (e.g.\ Table~\ref{tab:chaining-uniform} at $\alpha{\approx}122$); cheaper than trees if redistribution shortens bins. Treeify when length remains high after a sane $m$. \\
\bottomrule
\end{tabular}
\end{table}

\paragraph{Resize versus treeify.}
Rehashing spreads keys when the hash function is sound but the table is simply too small. On our uniform-hash microbenchmark with fixed $m{=}4096$, $\alpha{\approx}122$ already explains list avg.\ comparisons ${\sim}62$ and max chain 167---that is overloaded capacity, not Zipf skew, and a normal library would resize. Tree buckets address the complementary case: individual bins stay long even after a larger table---coarse routing keys, posting structures replayed as chains, or adversarial bucket ids. Our forced-bucket microbenchmark (Table~\ref{tab:chaining-skew}) isolates that second failure mode. Table~\ref{tab:scale-baseline} ($\alpha{=}16$) shows the middle ground where lists remain competitive while trees still cut comparisons.

\paragraph{Reproducibility.}
All timings and memory figures in Sections~\ref{sec:chaining-benchmark}--\ref{sec:library-benchmark} can be regenerated with \texttt{make chaining\_benchmark} and \texttt{scripts/bench\_repeat.sh}; see Appendix~\ref{app:methodology} and repository tag \texttt{v0.0.1}~\cite{rbtree_repo}.

\section{Conclusion}

Relative to list chaining, treeify policy (hybrid-batch vs hybrid-incremental) and load factor matter as much as the constant~$k$ for C implementers. Forced-bucket and posting-length evidence favor hybrid-incremental or always-tree when bins stay long; rehash when high $\alpha$ is only deferred resize. Stable metrics are comparisons and heap; long-list wall-clock is secondary. Source, Make targets, and table generators ship at repository tag \texttt{v0.0.1}~\cite{rbtree_repo}.

\section{Code and Experiment Availability}
All source code used for the experiments in this paper is published in the accompanying GitHub repository~\cite{rbtree_repo}. The repository is the reproducibility package: C implementations, benchmark drivers used for the tables in this article, and optional trie/quantile experiments omitted here for length.

The paper reports measured behaviour and design guidance; the repository provides the exact data-structure operations, key generation, and benchmark loops.

\paragraph{Data availability statement.}
All C sources, Make targets (\texttt{make run-chaining-compare}, \texttt{make run-chaining-uniform-repeated}, \texttt{make run-chaining-scale}, \texttt{scripts/bench\_repeat.sh}), and vendor word-list instructions are available at \url{https://github.com/rbtreechainingforhashtable/project} under Git tag \texttt{v0.0.1} (includes \texttt{heap\_bytes}, hybrid-batch/incremental policies, and the 2026-07-15 regenerated tables). Chaining benchmark rows in Tables~\ref{tab:chaining-uniform}--\ref{tab:variance-chaining} were regenerated on 2026-07-15 on the platform in Table~\ref{tab:platform}; long-running forced-bucket and application rows are single-run measurements as noted in Appendix~\ref{app:methodology}.

\paragraph{Conflicts of interest.}
The author declares no conflicts of interest.

\paragraph{Funding.}
This research received no specific grant from any funding agency in the public, commercial, or not-for-profit sectors.

\appendix
\section{Benchmark Methodology}
\label{app:methodology}

This appendix specifies the hardware, build, timing, datasets, metrics, and repetition policy used for the experiments in Section~\ref{sec:evaluation}. The same definitions are maintained in \texttt{docs/benchmark-methodology.md} in the reproducibility repository~\cite{rbtree_repo}.

\subsection{Hardware and software platform}

All timings reported in this paper were collected on a single workstation:

\begin{table}[H]
\centering
\caption{Measurement platform}
\label{tab:platform}
\begin{tabular}{l l}
\toprule
Component & Configuration \\
\midrule
CPU & Apple M4 Pro \\
RAM & 48~GB \\
Operating system & macOS 26.5.1 (build 25F80) \\
C compiler & Apple Clang 16.0.0 (\texttt{cc}, Xcode command-line tools) \\
C library & Apple libc (system) \\
\bottomrule
\end{tabular}
\end{table}

The machine was otherwise idle during timed phases. Hypervisor or cloud instances were not used.

\subsection{Build configuration}

Every experiment binary is built from the repository Makefile with default settings unless noted otherwise:

\begin{itemize}
\item \texttt{CC=cc}
\item \texttt{CFLAGS=-O2 -Wall -Wextra}
\item link-time optimization (LTO) and profile-guided optimization (PGO) \emph{disabled}
\item \texttt{hashtable\_benchmark} additionally links vendored TommyDS~\cite{tommyds} with \texttt{-DWITH\_TOMMY}
\end{itemize}

Reproduction command: \texttt{make clean \&\& make}. The exact compile line for any target can be inspected with \texttt{make -n TARGET}.

\subsection{Timing methodology}

All drivers call \texttt{clock\_gettime(CLOCK\_MONOTONIC, ...)} through \texttt{common/timing.h}. This measures \textbf{elapsed wall time} between two monotonic timestamps:

\begin{itemize}
\item \textbf{Included:} time waiting in the run queue, page faults, heap allocation, and I/O while the timed phase executes.
\item \textbf{Not used:} thread CPU time (\texttt{CLOCK\_THREAD\_CPUTIME\_ID}), \texttt{getrusage} user/system time, or Java-style JVM warm-up iterations.
\end{itemize}

Each benchmark defines one or more \emph{phases} (Table~\ref{tab:bench-phases}). A timer starts immediately before the phase body and stops immediately after it completes. Argument parsing, opening input files, and allocating empty containers are excluded unless the driver README states otherwise.

\begin{table*}[!t]
\centering
\footnotesize
\caption{Timed phases by experiment driver (monotonic wall clock per phase).}
\label{tab:bench-phases}
\begin{tabular}{@{}lll@{}}
\toprule
Driver & Insert phase & Query phase \\
\midrule
\texttt{chaining\_benchmark} & insert all keys & lookup every key \\
\texttt{inverted\_chain\_bench} & build trigram index / replay lengths & chain lookup replay \\
\texttt{hashtable\_benchmark} & generate, insert (, optional file I/O) & lookup each key \\
\bottomrule
\end{tabular}
\end{table*}

\subsection{Datasets and workload sizes}

Table~\ref{tab:datasets} lists inputs. Synthetic workloads fix the pseudorandom seed (\texttt{--seed 1} by default) so key sequences repeat across runs.

\begin{table}[H]
\centering
\caption{Datasets and default workload parameters}
\label{tab:datasets}
\begin{tabularx}{\linewidth}{@{}
  >{\raggedright\arraybackslash}p{0.26\linewidth}
  >{\raggedright\arraybackslash}X
  >{\raggedright\arraybackslash}X
  @{}}
\toprule
Experiment & Input & Default size \\
\midrule
Trigram posting CDF / replay
  & {\footnotesize\nolinkurl{vendor/english-words/words_alpha.txt}~\cite{words_alpha}}
  & 370{,}105 words; hottest 256--2048 lists replayed \\
Same-API scale baseline
  & PRNG alphanumeric strings
  & $2^{20}$ keys, $2^{16}$ buckets, seed~1 \\
External library context
  & PRNG alphanumeric strings
  & $2^{25}$ keys (optional; includes file I/O) \\
Chaining microbenchmark
  & uniform-hash FNV (fixed~$m$) or forced-bucket
  & $5\times10^5$ keys, 4096 buckets ($\alpha{\approx}122$), or 8~hot bins (stress) \\
\bottomrule
\end{tabularx}
\end{table}

Forced-bucket stress keys have the form \texttt{BBBB:seq:suffix}; the bucket hash returns \texttt{BBBB}, concentrating roughly $N/\text{hot\_buckets}$ keys per overloaded bin. This construction is a per-bucket chaining probe (hash quality held out), not an emulation of Zipf hash collisions across thousands of bins. The uniform-hash microbenchmark uses FNV into a fixed $m{=}4096$ without resize, so $\alpha{\approx}122$ deliberately overloads capacity; interpret tree wins there alongside the resize-versus-treeify discussion, not as ``short-chain uniform.''

\paragraph{Threats to validity (workload framing).}
The equal eight-bin stress maximizes symmetry and wall-clock exposure of list/hybrid-batch costs; it does not reproduce naturalistic bucket occupancy histograms under FNV. The fixed-$m$ uniform-hash run ($\alpha{\approx}122$) overstates how often trees beat lists in a resizing production table---Table~\ref{tab:scale-baseline} at $\alpha{=}16$ is the fairer moderate-load baseline. Zipf over hashed $n$-gram keys is out of scope: Layer~B already supplies empirical length realism via the posting CDF and replay. Results attribute insert/lookup differences to overflow structure and treeify policy, not to hash-function design.

\paragraph{Threats to validity (Java protocol).}
Hybrid-incremental approximates \texttt{HashMap}'s convert-as-you-go timing under our insert/mid-load mix; it is not a binary-compatible port (no untreeify at six nodes, no JVM tree-bin packaging, no shared resize path). Hybrid-batch is deliberately the non-Java cargo-cult baseline. Guidelines that cite ``hybrid'' without batch/incremental would be shaky; ours name both.

\subsection{Metric definitions}

\begin{table}[H]
\centering
\caption{Reported metrics}
\label{tab:metrics}
\begin{tabular}{p{0.24\textwidth} p{0.68\textwidth}}
\toprule
Metric & Definition \\
\midrule
\texttt{insert\_seconds}, \texttt{init\_seconds}, \texttt{load\_seconds} & Monotonic elapsed seconds for the build/insert phase (Table~\ref{tab:bench-phases}) \\
\texttt{search\_seconds} & Monotonic elapsed seconds for the query phase \\
\texttt{comparisons} & Number of \texttt{strcmp} calls during chained-hash lookup (\texttt{chain\_ht}) \\
\texttt{avg\_comparisons} & \texttt{comparisons} / number of successful lookups \\
\texttt{max\_bucket} & Maximum entries in any hash bucket after inserts \\
\texttt{heap\_bytes} & Heap footprint after inserts (\texttt{chain\_ht\_heap\_bytes()}) \\
\texttt{treeify\_events} & Number of list$\to$tree conversions (hybrid modes) \\
\texttt{memory} & TommyDS heap usage (bytes); external-library context only \\
\bottomrule
\end{tabular}
\end{table}

Drivers emit one machine-readable \texttt{experiment=...} line on \textbf{stderr} per run; \texttt{scripts/bench\_repeat.sh} aggregates repeated runs.

\subsection{Number of runs and variance}

Uniform chaining configurations were repeated \textbf{five} times with \texttt{make run-chaining-uniform-repeated} (uniform workload only---no skew runs between repetitions). Table~\ref{tab:variance-chaining} reports mean~$\pm$~sample standard deviation (Bessel correction, $n{-}1$ denominator). Deterministic counters (\texttt{avg\_comparisons}, \texttt{max\_bucket}) are identical across runs with the same seed.

\begin{table}[H]
\centering
\caption{Five-run variance, uniform-only chaining ($5\times10^5$ keys, 4096 buckets, seed~1). Insert and lookup: wall time (s, mean~$\pm$~sample stddev).}
\label{tab:variance-chaining}
\begin{tabular}{@{}lrr@{}}
\toprule
Policy & Insert (s) & Lookup (s) \\
\midrule
List & $1.073 \pm 0.065$ & $1.003 \pm 0.043$ \\
Hybrid-batch ($k{=}8$) & $1.152 \pm 0.035$ & $0.059 \pm 0.003$ \\
Hybrid-incremental ($k{=}8$) & $0.166 \pm 0.003$ & $0.055 \pm 0.002$ \\
Always RB tree & $0.160 \pm 0.005$ & $0.053 \pm 0.001$ \\
\bottomrule
\end{tabular}
\end{table}

Earlier full-suite repeats (skew interleaved with uniform) inflated uniform stddevs; those are not used for Table~\ref{tab:variance-chaining}. Long-running forced-bucket list/batch rows (Tables~\ref{tab:chaining-policy}, \ref{tab:chaining-skew}) and the external-library context table remain single-run. List wall-clock under overload is especially noise-sensitive (repeat runs on the same seed have differed by more than $2\times$); tree/hybrid lookup times and all comparison counters are stable. Claims that depend on absolute list seconds should not be treated as primary results. The in-memory scale baseline (Table~\ref{tab:scale-baseline}) is a single run regenerated with \texttt{make run-chaining-scale}.

\section{Experiment Reproducibility Notes}
The experiments reported in this paper are tied to the repository artifacts as follows:
\begin{itemize}
\item The trigram posting CDF and chain replay use \texttt{inverted\_chain\_bench} over \texttt{words\_alpha.txt}.
\item The list/tree/hybrid chaining measurements use \texttt{chaining\_benchmark} (\texttt{compare}, \texttt{policy}, \texttt{scale} suites) with the shared \texttt{chain\_ht} implementation.
\item Uniform variance uses \texttt{make run-chaining-uniform-repeated}; in-memory scale uses \texttt{make run-chaining-scale}.
\item An optional external-library context benchmark compares the always-tree table with GNU \texttt{hsearch} and \texttt{tommy\_hashdyn} under a generated-key workload that includes file I/O.
\item Optional trie and quantile drivers remain in the repository for readers who want those case studies.
\end{itemize}

For reproducibility, the most important parameters are the hash-table size, the hash function, the key distribution, the treeify policy/threshold, and the number of inserted elements: they control collision frequency, chain length, and measured list-versus-tree differences.

\section{Interpreting Benchmark Results}
The benchmark results should be interpreted as implementation-level measurements on the platform and build described in Appendix~\ref{app:methodology}. Absolute timings change with CPU model, compiler version, optimization flags, cache behavior, and operating-system scheduling; five-run uniform chaining tests (Table~\ref{tab:variance-chaining}) show typical run-to-run spread for insert and lookup phases.

When the hash function distributes keys uniformly, linked-list chaining remains competitive because most chains stay short and the extra balancing work of an RB tree may not be compensated by faster lookup. When distribution is uneven, the tree bounds the per-bucket search cost more effectively than a list and can reduce the worst-case effect of long chains.

\end{document}